\documentclass[final]{aipproc}
  
\usepackage{subfig}
\pdfoutput=1
\layoutstyle{8x11single}

\newcommand{\germanium}{$^{76}$Ge~}
\newcommand{\germaniumsp}{$^{76}$Ge}

\newcommand{\nbbsp}{$0\nu\beta\beta$}
\newcommand{\nbb}{$0\nu\beta\beta$~}
\newcommand{\mbb}{$\langle m_{\beta\beta} \rangle$~}
\newcommand{\mjd}{{\sc Majorana Demonstrator}~}

\newcommand{\demonstrator}{{\sc Demonstrator}~}
\newcommand{\demonstratorsp}{{\sc Demonstrator}}

\makeatletter
\let\@fnsymbol=\@arabic
\makeatother

\begin{document}

\title{Status of the {\sc Majorana Demonstrator} experiment}

\classification{N/A}
\keywords{Neutrinoless double beta-decay, germanium, PPC}

\newcommand{\alberta}{Centre for Particle Physics, University of Alberta, Edmonton, AB, Canada}
\newcommand{\blhill}{Department of Physics, Black Hills State University, Spearfish, SD, USA}
\newcommand{\ITEP}{Institute for Theoretical and Experimental Physics, Moscow, Russia}
\newcommand{\JINR}{Joint Institute for Nuclear Research, Dubna, Russia}
\newcommand{\lbnl}{Nuclear Science Division, Lawrence Berkeley National Laboratory, Berkeley, CA, USA}
\newcommand{\lanl}{Los Alamos National Laboratory, Los Alamos, NM, USA}
\newcommand{\uw}{Center for Experimental Nuclear Physics and Astrophysics, 
and Department of Physics, University of Washington, Seattle, WA, USA}
\newcommand{\uchic}{Department of Physics, University of Chicago, Chicago, IL, USA}
\newcommand{\unc}{Department of Physics and Astronomy, University of North Carolina, Chapel Hill, NC, USA}
\newcommand{\duke}{Department of Physics, Duke University, Durham, NC, USA}
\newcommand{\ncsu}{Department of Physics, North Carolina State University, Raleigh, NC, USA}
\newcommand{\ornl}{Oak Ridge National Laboratory, Oak Ridge, TN, USA}
\newcommand{\ou}{Research Center for Nuclear Physics and Department of Physics, Osaka University, Ibaraki, Osaka, Japan}
\newcommand{\pnnl}{Pacific Northwest National Laboratory, Richland, WA, USA}
\newcommand{\ttu}{Tennessee Tech University, Cookeville, TN, USA}
\newcommand{\sdsmt}{South Dakota School of Mines and Technology, Rapid City, SD, USA}
\newcommand{\sjtu}{Shanghai Jiaotong University, Shanghai, China}
\newcommand{\usc}{Department of Physics and Astronomy, University of South Carolina, Columbia, SC, USA}
\newcommand{\usd}{Department of Physics, University of South Dakota, Vermillion, SD, USA} 
\newcommand{\ut}{Department of Physics and Astronomy, University of Tennessee, Knoxville, TN, USA}
\newcommand{\tunl}{Triangle Universities Nuclear Laboratory, Durham, NC, USA}

\author{R.D.~Martin}{address={\lbnl},altaddress={\usd}, email={ryan.martin@usd.edu}}
\author{N.~Abgrall}{address={\lbnl}	}	
\author{E.~Aguayo}{address={\pnnl} }
\author{F.T.~Avignone~III}{address={\usc},altaddress={\ornl}}
\author{A.S.~Barabash}{address={\ITEP}}
\author{F.E.~Bertrand}{address={\ornl}}
\author{M.~Boswell}{address={\lanl}}
\author{V.~Brudanin}{address={\JINR}}
\author{M.~Busch}{address={\duke},altaddress={\tunl}}
\author{A.S.~Caldwell}{address={\sdsmt}}
\author{Y-D.~Chan}{address={\lbnl}}
\author{C.D.~Christofferson}{address={\sdsmt}}
\author{D.C.~Combs}{address={\ncsu},altaddress={\tunl}} 
\author{J.A.~Detwiler}{address={\uw}}	
\author{P.J.~Doe}{address={\uw}} 
\author{Yu.~Efremenko}{address={\ut}}
\author{V.~Egorov}{address={\JINR}}
\author{H.~Ejiri}{address={\ou}}
\author{S.R.~Elliott}{address={\lanl}}
\author{J.~Esterline}{address={\duke},altaddress={\tunl}}
\author{J.E.~Fast}{address={\pnnl}}
\author{P.~Finnerty}{address={\unc},altaddress={\tunl}}
\author{F.M.~Fraenkle}{address={\unc},altaddress={\tunl}} 
\author{A.~Galindo-Uribarri}{address={\ornl}}	
\author{G.K.~Giovanetti}{address={\unc},altaddress={\tunl}}  
\author{J. Goett}{address={\lanl}}	
\author{M.P.~Green}{address={\unc},altaddress={\tunl}}  
\author{J.~Gruszko}{address={\uw}}
\author{V.E.~Guiseppe}{address={\usc}}
\author{K.~Gusev}{address={\JINR}}
\author{A.L.~Hallin}{address={\alberta}}
\author{R.~Hazama}{address={\ou}}
\author{A.~Hegai}{address={\lbnl}}
\author{R.~Henning}{address={\unc},altaddress={\tunl}}
\author{E.W.~Hoppe}{address={\pnnl}}
\author{S.~Howard}{address={\sdsmt}}
\author{M.A.~Howe}{address={\unc},altaddress={\tunl}}
\author{K.J.~Keeter}{address={\blhill}}
\author{M.F.~Kidd}{address={\ttu}}
\author{O.~Kochetov}{address={\JINR}}
\author{S.I.~Konovalov}{address={\ITEP}}
\author{R.T.~Kouzes}{address={\pnnl}}
\author{B.D.~LaFerriere}{address={\pnnl}} 
\author{J.~Leon}{address={\uw}}
\author{L.E.~Leviner}{address={\ncsu},altaddress={\tunl}}
\author{J.C.~Loach}{address={\sjtu},altaddress={\lbnl}}
\author{J.~MacMullin}{address={\unc},altaddress={\tunl}}
\author{S.~MacMullin}{address={\unc},altaddress={\tunl}}
\author{S.~Mertens}{address={\lbnl}}
\author{L.~Mizouni}{address={\usc},altaddress={\pnnl}}
\author{M.~Nomachi}{address={\ou}}
\author{J.L.~Orrell}{address={\pnnl}}
\author{C. O'Shaughnessy}{address={\unc},altaddress={\tunl}}	
\author{N.R.~Overman}{address={\pnnl} }
\author{D.G.~Phillips II}{address={\ncsu},altaddress={\tunl}}  
\author{A.W.P.~Poon}{address={\lbnl}}
\author{K.~Pushkin}{address={\usd}}
\author{D.C.~Radford}{address={\ornl}}
\author{K.~Rielage}{address={\lanl}}
\author{R.G.H.~Robertson}{address={\uw}}
\author{E.~Romero-Romero}{address={\ut},altaddress={\ornl}}
\author{M.C.~Ronquest}{address={\lanl}}
\author{A.G.~Schubert}{address={\uw}}
\author{B.~Shanks}{address={\unc},altaddress={\tunl}}	
\author{T.~Shima}{address={\ou}}
\author{M.~Shirchenko}{address={\JINR}}
\author{K.J.~Snavely}{address={\unc},altaddress={\tunl}}
\author{N.~Snyder}{address={\usd}}
\author{A.~Soin}{address={\pnnl}}
\author{A.M.~Suriano}{address={\sdsmt}}
\author{J. Thompson}{address={\blhill},altaddress={\sdsmt}}
\author{V.~Timkin}{address={\JINR}}
\author{W.~Tornow}{address={\duke},altaddress={\tunl}}
\author{R.L.~Varner}{address={\ornl}}
\author{S.~Vasilyev}{address={\ut}}
\author{K.~Vetter}{address={\lbnl}}
\author{K.~Vorren}{address={\unc},altaddress={\tunl}} 
\author{B.R.~White}{address={\ornl}}
\author{J.F.~Wilkerson}{address={\unc},altaddress={\tunl, and \ornl}}
\author{W.~Xu}{address={\lanl}}
\author{E.~Yakushev}{address={\JINR}}
\author{A.R.~Young}{address={\ncsu},altaddress={\tunl}}
\author{C.-H.~Yu}{address={\ornl}}
\author{V.~Yumatov}{address={\ITEP}}


\begin{abstract}
The {\sc Majorana Demonstrator} neutrinoless double beta-decay experiment is currently under construction at the Sanford Underground Research Facility in South Dakota, USA. An overview and status of the experiment are given.
\end{abstract}

\maketitle


\section{Neutrinoless double beta-decay}
Searches for neutrinoless double-beta (\nbbsp) decay are the most promising way to determine whether neutrinos are Majorana or Dirac particles \cite{Zralek:1997sa}. Discovering that neutrinos are Majorana particles would have a profound impact on our understanding of matter \cite{Schwingenheuer:2012zs,Bilenky:2012qi}. The observation of the decay would show that lepton number is violated providing a mechanism for leptogenesis to explain the matter anti-matter asymmetry in the Universe \cite{Fukugita:1986hr}. Furthermore, the Majorana nature of the neutrino would allow for the see-saw mechanism \cite{GellMan:1979, PhysRevLett.44.912} to explain the, seemingly finely-tuned, small neutrino masses. Finally, the rate of  neutrinoless double beta-decay could be used to determine the neutrino mass scale \cite{Vergados:2012xy}.

In the standard paradigm, neutrinoless double beta-decay is described by the simultaneous decay of two neutrons into two protons and two electrons, mediated by the exchange of two W bosons and a virtual Majorana neutrino. It has however been shown that any (non-standard) mechanism that allows for neutrinoless double beta-decay would imply that neutrinos are Majorana particles \cite{Hirsch:2006yk}. Extracting information about neutrinos, such as their mass scale, from the observation of \nbb decay is complicated by several factors. Even in the standard paradigm, uncertainties in the nuclear matrix elements make it difficult to use the decay rate to precisely determine the neutrino mass scale. Furthermore, it is not yet understood if the axial vector current coupling constant, $g_A$, is effectively reduced in \nbb decay as it is in the two neutrino double beta-decay and single beta-decay \cite{Vogel:2012ja,Kotila:2012zza}. For a given neutrino mass and nuclear matrix element, $g_A$ could change the decay rate by a factor of around 20 \cite{Robertson:2013cy}. Keeping these limitation in mind, the lifetime of \nbb decay, $T_{1/2}^{0\nu}$, in the standard paradigm, can be written as \cite{Vogel:2012ja}:
\begin{equation}
T_{1/2}^{0\nu}=\left(G_{0\nu} |M_{0\nu}|^2 \left(\frac{\langle m_{\beta\beta} \rangle}{m_e}\right)^2  \right)^{-1} \\
\end{equation}
where, $G_{0\nu}$, is the phase-space factor, $M_{0\nu}$, is the nuclear matrix element, $m_e$, is the mass of the electron and, $\langle m_{\beta\beta} \rangle$, is the effective Majorana neutrino mass. The effective neutrino mass, resulting from the mixing of neutrino eigenstates that can participate in the decay, is given by:
\begin{equation}
\langle m_{\beta\beta} \rangle = \left| \sum_{i=1}^3 U_{ei}^2 m_i \right|\\
\end{equation}
where $m_i$ are the masses of the neutrino energy eigenstates and, $U_{ei}$, are the elements of the neutrino mixing matrix.

The best current experimental limits on the lifetime for the decay indicate a half life longer than 3.0$\times 10^{25}$ years in \germanium \cite{Agostini:2013mzu} and 3.4$\times 10^{25}$ in $^{136}$Xe \cite{Gando:2012zm}. These limits translate to an upper limit on \mbb of order $\sim$ 200\,meV, depending on the choice of nuclear matrix element and choice of $g_A$. The current generation of experiments are expected to reach sensitivities in the range of 100\,meV. The focus for the next generation of experiments will be to scan the inverted hierarchy region of the possible \nbb decay, which corresponds to lowering the sensitivity by roughly one order of magnitude (\mbb$\sim$10\,meV). This range is appealing as it corresponds to \mbb in the range of the atmospheric neutrino mixing mass-squared difference. These next generation experiments will require fiducial masses of the order of 1 tonne and exposures of several years to reach this sensitivity. It has also been argued that no candidate isotope is a particularly better choice for an experiment \cite{Robertson:2013cy}. However, for a signal to be convincing, it is generally agreed that it should be observed in several isotopes. It is thus critical for several large neutrinoless double beta-decay experiments to be developed.

\section{Experimental searches for neutrinoless double beta-decay}
Experiments to search for \nbb decay are designed to precisely measure the energy spectrum of the two electrons emitted in the decay. Experiments are optimized to be sensitive near the Q-value of the decay, where a small peak in the electron energy spectrum would reveal the neutrinoless mode of the decay. The sensitivity of \nbb experiments is limited by the Poisson statistics in a region of interest near the expected peak with a width related to the energy resolution. In the presence of backgrounds, the 1$\sigma$ sensitivity in the detectable half life from a target of mass, $M$, observed during a time, $t$, can be written \cite{Cremonesi:2012av}:
\begin{equation}
T_{1/2}^{0\nu}=\ln{2} \frac{\epsilon a x N_A}{A} \sqrt{\frac{Mt}{\sigma_E B}}\\
\label{eqn:sensitivity}
\end{equation}
where, $\epsilon$, is the efficiency of the experiment at detecting the decay, $a$, is the isotopic abundance of the target, $N_A$, is Avogadro's Number, x, is the number of atoms per molecule that can undergo double beta-decay, $A$, is the molecular mass of the target,  $\sigma_E$, is the energy resolution of the experiment and, $B$, is the background rate in the region of interest (per unit target mass and per energy, e.g. counts/keV/kg/yr). Increasing the isotopic abundance and the efficiency of detecting the decay have the biggest effect in increasing the sensitivity and are typically set by the choice of technology. The achievable energy resolution is also dictated by the choice of technology. The $Mt$ ``exposure'' term is dependent upon the size of the experiment. Finally, the ultimate limit for a given technology and size of experiment will be achieved by minimizing the background rate in the region of interest. Thus, once a technology choice has been made, experimental designs focus on achieving the lowest possible background rate so that larger exposures remain competitive.

\section{P-type point contact HPGe detectors}
P-type point contact (PPC) high purity germanium (HPGe) detectors \cite{luke_89, Barbeau:2007qi} are a promising technology choice to search for neutrinoless double beta-decay. HPGe detectors are a well established technology with excellent energy resolution (approximately 0.2\% at 2039\,keV, the Q-value for the \nbb decay of \germaniumsp). The detectors can be made from germanium crystals enriched in the \nbbsp -candidate isotope \germaniumsp, thus optimizing the abundance and efficiency terms in Eqn. \ref{eqn:sensitivity}. In addition, the point contact detector geometry (see Fig. \ref{fig:mse1332}a) gives the ability to distinguish between single and multiple site energy depositions in the crystal \cite{Barbeau:2007qi,Budjas:2009zu,Cooper2011303}. This is particularly useful in identifying the backgrounds from gamma rays scattering multiple times in a crystal and distinguishing them from the localized energy deposition of the two electrons from a double beta-decay. Fig. \ref{fig:mse1332}b shows the recorded trace of the charge as a function of time (black solid line) from a PPC detector for a 1332\,keV gamma ray scattering four times within the crystal. The red (dashed) line shows the corresponding current in the point contact detector (obtained by differentiating the charge trace) where four different current pulses corresponding to the four scatters can easily be seen. Through pulse shape discrimination, the backgrounds from multiple scatter events in PPCs can be greatly reduced.

\begin{ltxfigure}
\centering
  \subfloat[PPC detector]{\includegraphics[height=175pt]{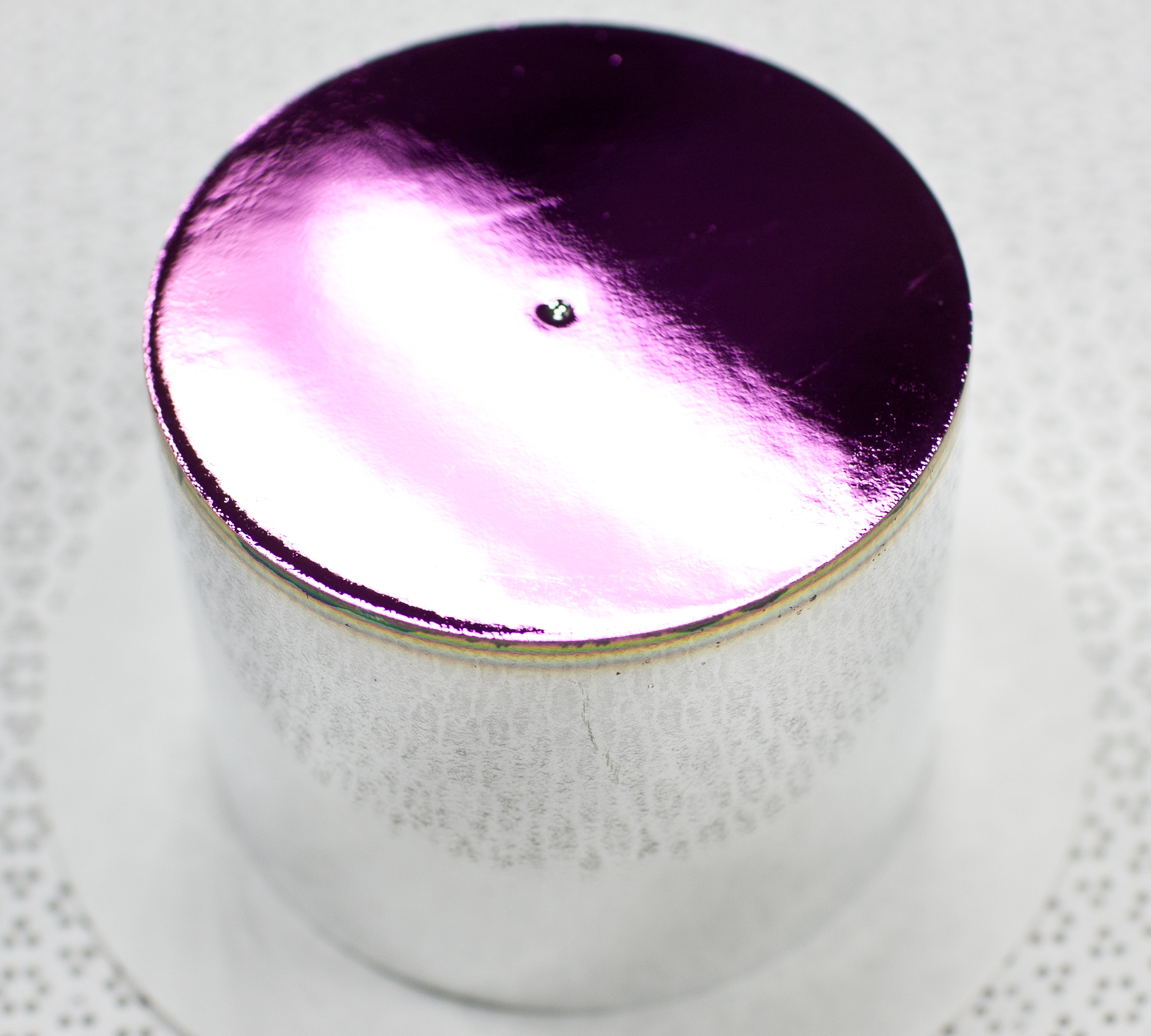}}
  \subfloat[Charge and current signals]{\includegraphics[height=175pt]{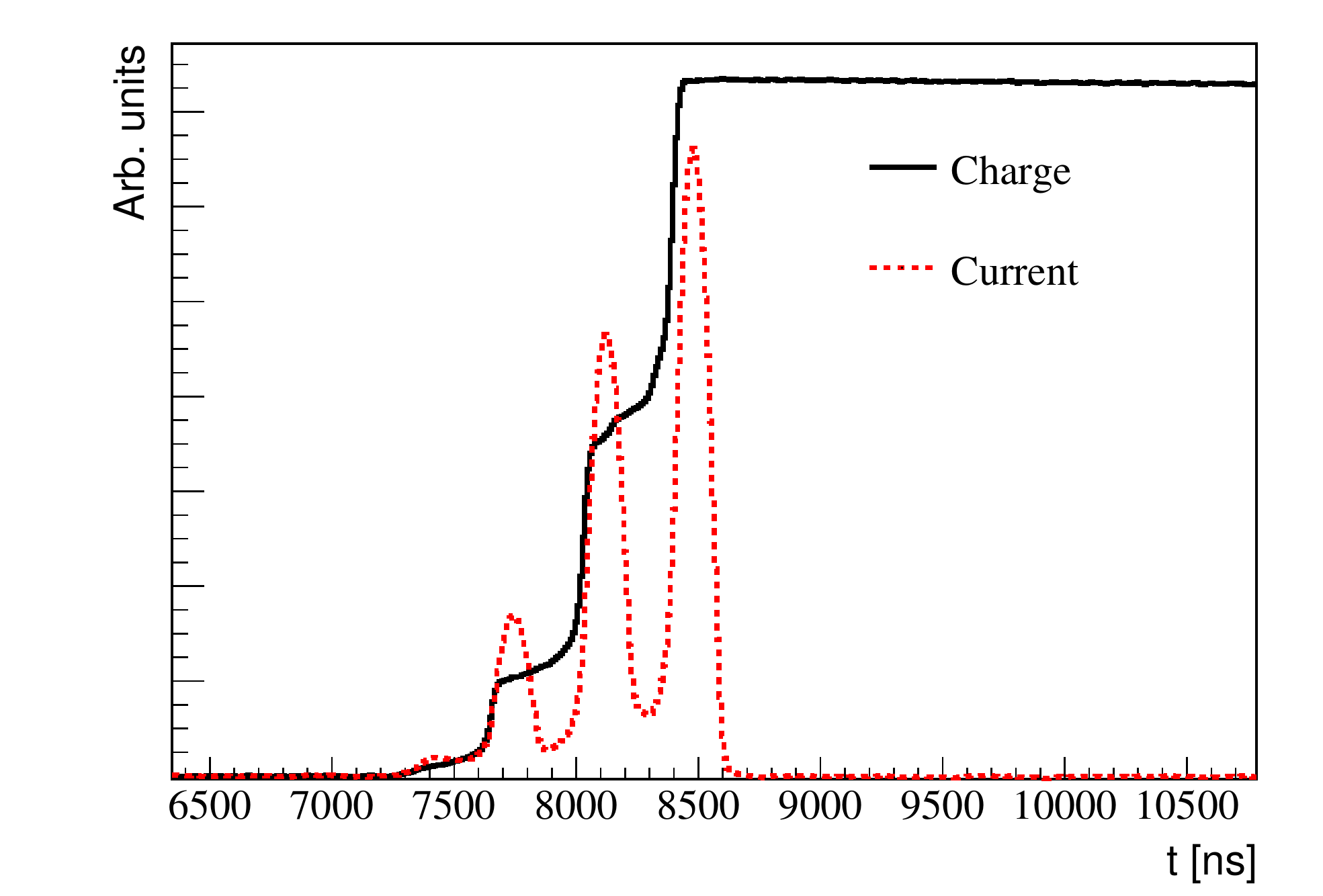}}
  \caption{\label{fig:mse1332}Left panel: PPC detector that was fabricated at Lawrence Berkeley National Laboratory (LBNL) and described in \cite{Martin:2011vj,Aguayo:2012up}. Right panel: charge (black solid) and current (red dotted) as a function of time recorded with the LBNL PPC. The current was obtained by differentiating the charge trace from this 1332\,keV gamma ray event \cite{Martin:2011vj}. At least four different current pulses can be distinguished, corresponding to four different interactions (Compton scatters) of the gamma ray in the detector. The ability to identify multiple scatter events gives PPC detectors the capability of rejecting backgrounds to the \nbb decay signal.}
\end{ltxfigure}

\section{The {\sc Majorana Demonstrator}}
\subsection{Overview}
\begin{figure}
\centering
  \includegraphics[width=0.7\textwidth]{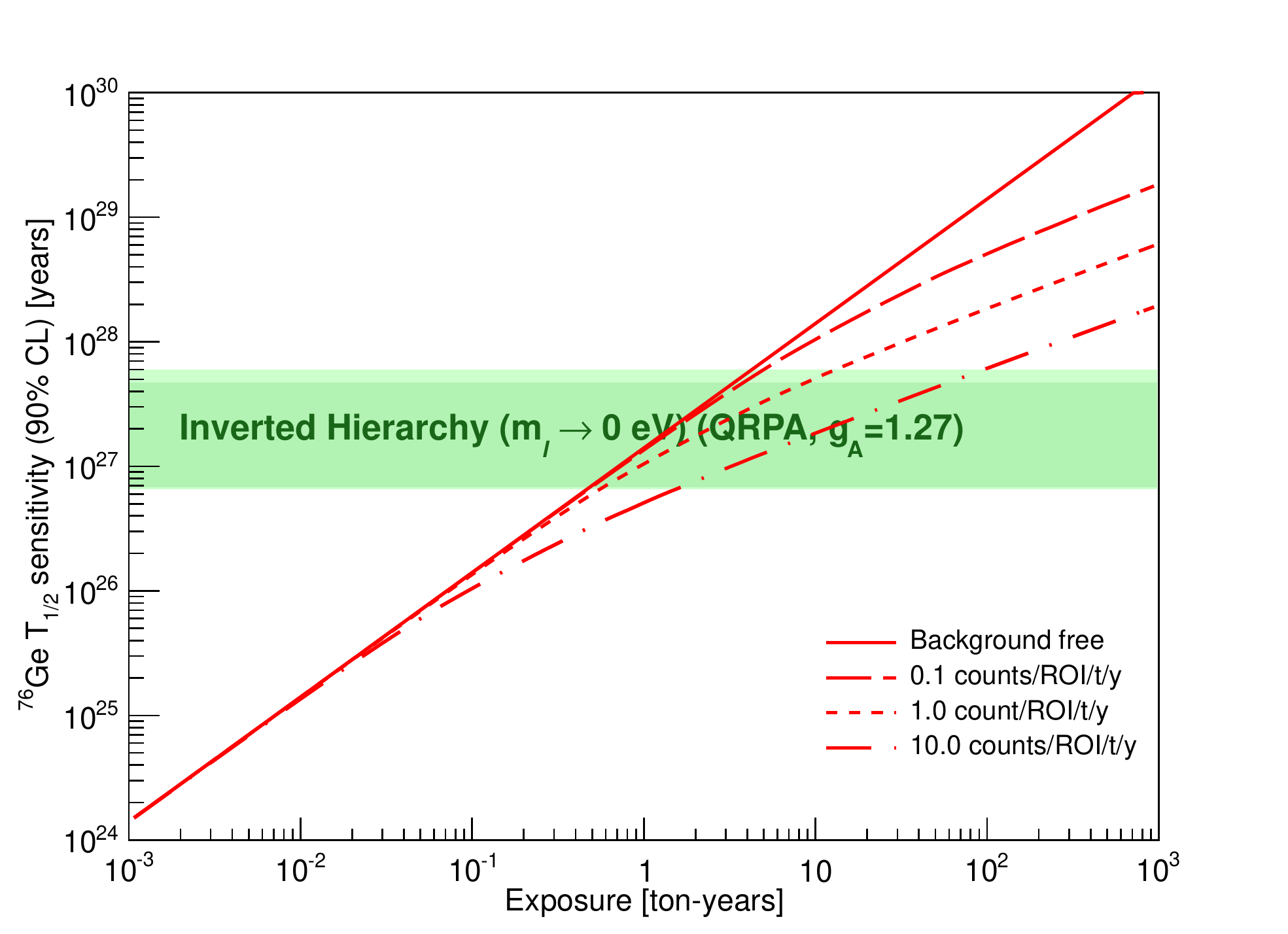}
  \caption{\label{fig:sensitivity}Sensitivity of a germanium-based experiment such as the \mjd . The sensitivity to a 90\% confidence measurement of the half-life as a function of the exposure is calculated using the Feldman-Cousins method \cite{FeldmanCousins:1997}. The band shows the region of the inverted hierarchy, with the nuclear matrix element calculated using the Quasi-Random Particle Approximation \cite{Faessler:2013hz} and a nominal value of $g_A$. The effect of different background rates in a 4\,keV the region of interest (ROI) is shown, highlighting the need for a very low background rate. The \mjd will have an exposure of the order of 60\,kgy after two years. It can be seen that in order to probe the inverted hierarchy, tonne scale experiments are necessary.}
\end{figure}

The \mjd experiment \cite{mjahep:2013} is deploying an array of 40\,kg of PPC detectors, of which, up to 30\,kg will be enriched to at least 86\% in \germaniumsp. The experiment is currently under construction in the Davis Campus at the 4850\,ft level of the Sanford Underground Research Facility (SURF), in South Dakota, USA. The primary goal is to demonstrate a technology that can be scaled up to a tonne scale experiment to perform a search for \nbb decay in the inverted mass hierarchy region. The \demonstrator will also have the sensitivity to test the claim in \cite{KlapdorKleingrothaus:2006ff} as well as the ability to search for other rare events (see below). The sensitivity to a half-life measurement (see Eqn. \ref{eqn:sensitivity}) as a function of exposure is shown in Fig. \ref{fig:sensitivity} for germanium experiments. It can be seen that exposures of the order of tonne years are required to probe the inverted hierarchy regime. The \mjd thus aims to demonstrate a design, that when scaled up to 1 tonne, results in a background rate of 1 count per tonne of germanium per year in a 4\,keV region of interest near the Q-value.

The \mjd experiment will employ a conventional design for operating PPC detectors in vacuum cryostats surrounded by a compact shield. Extreme care is taken by using ultra-pure materials to minimize background from radioactive contaminants. The components near the detectors are fabricated from copper that is electroformed underground. Electroforming the copper increases the purity of the material (removing uranium and thorium) \cite{hoppe:2009}, while performing this operation underground significantly reduces the amount of cosmogenically produced isotopes in copper (such as $^{60}$Co). The collaboration has gone through substantial efforts to build a clean machine shop underground to fabricate components from the electroformed copper.

Each point contact detector is assembled in a low background mount fabricated from the electroformed copper and PTFE parts. The mount also holds a low mass front end board (LMFE) \cite{LMFE_paper_2011} used to read out the charge signal in the detector. The LMFE is part of a low background and low electronic noise readout system for the detector signals. The LMFE is assembled on a thin fused-silica wafer, with sputtered gold traces (on top of a sputtered titanium bonding layer) and a bare-die JFET to amplify the charge signals from the detectors. The readout system also includes low mass coaxial cables (50 Ohm impedance) with a diameter of approximately 1\,mm, fabricated from copper and FEP (fluorinated ethylene propylene) dielectric. These are attached to the LMFE using low background silver epoxy. The assembled mount is called a detector unit, which is shown in Fig. \ref{fig:DUandString}a. Detector units can be stacked into strings holding up to five units, as shown in Fig. \ref{fig:DUandString}b.

\begin{figure}
\centering
  \includegraphics[width=0.7\textwidth]{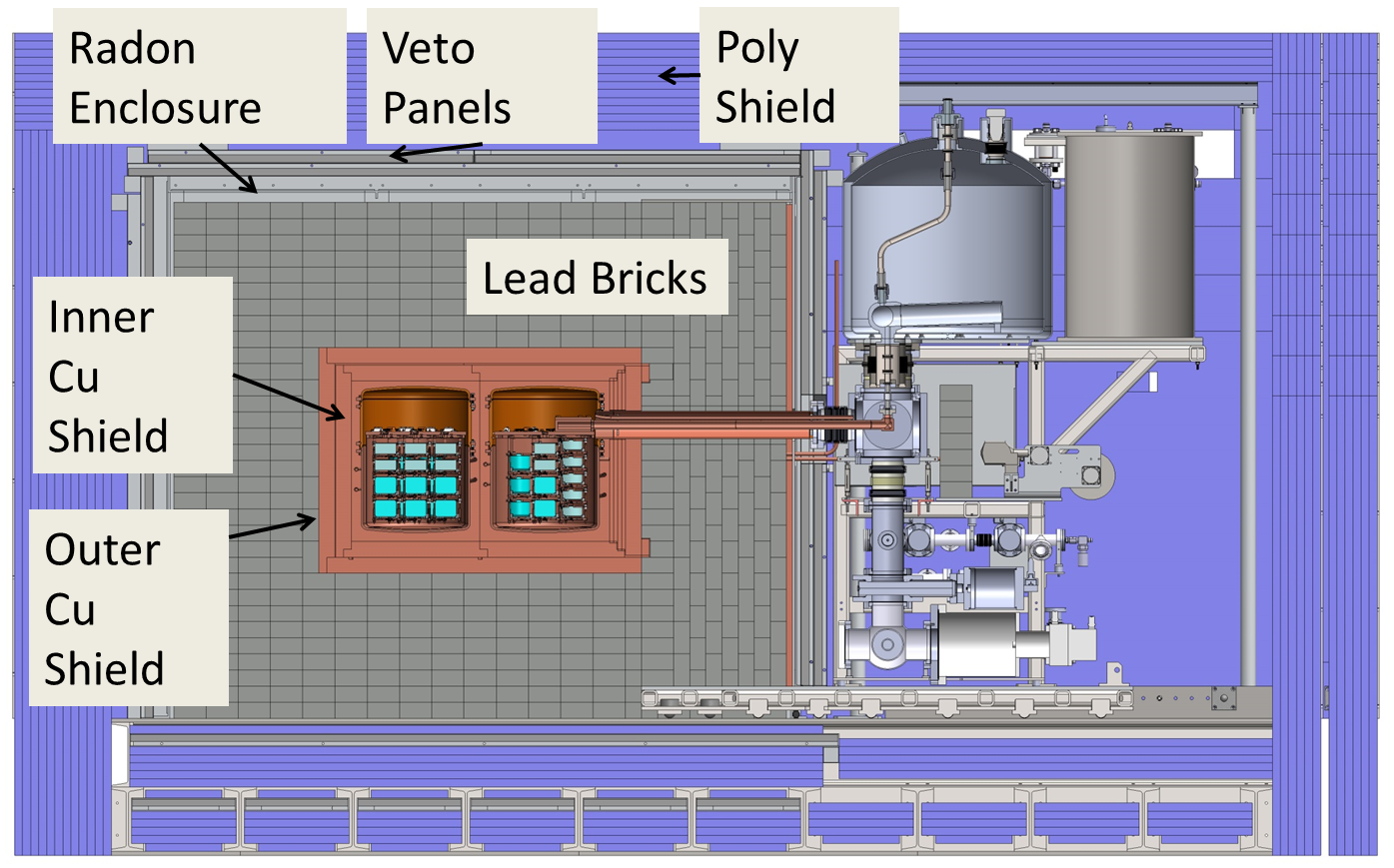}
  \caption{\label{fig:shield}Schematic of the \mjd experiment shield surrounding the two module cryostats. The vacuum and cryogenic systems for one of the cryostats is shown on the right. }
\end{figure}

The strings of detectors are then mounted into ``module'' vacuum cryostats that can hold up to seven strings. The module cryostats are cooled by a thermosyphon which circulates a closed loop of nitrogen \cite{Aguayo:2013ifa}. The nitrogen is condensed on one side of the circuit in a custom dewar and then evaporated on a cold plate near the detectors to provide the cooling power (see dewar on the far right of Fig. \ref{fig:shield}). This configuration has been shown to provide adequate cooling power while meeting the background goals for the experiment.

\begin{ltxfigure}[ht]
\centering
  \subfloat[Detector unit]{\includegraphics[width=0.45\textwidth]{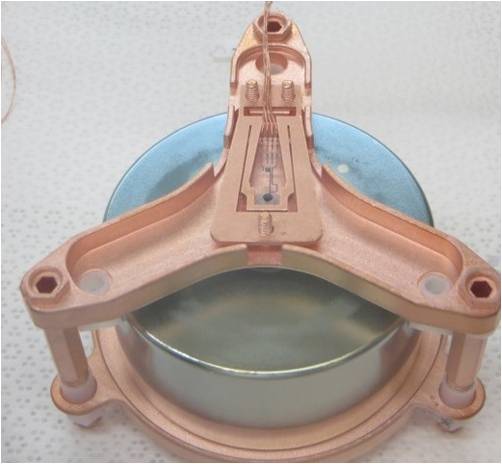}}  
  \subfloat[String of detector units]{\includegraphics[width=0.45\textwidth]{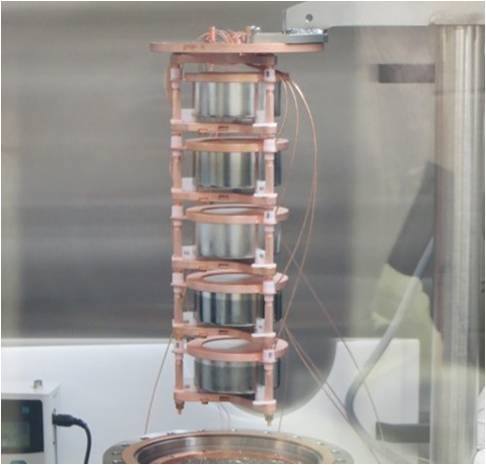}}    
  \caption{\label{fig:DUandString}Assembled detector unit (panel a), showing the germanium crystal in a copper mount with the low mass front end board on top to read out the charge signals. Panel b shows a string of five detetor units.}
\end{ltxfigure}

The \mjd experiment will have two module cryostats fabricated out of electroformed copper. The cryostats will be enclosed in a compact shield composed of several layers. From the inside out, these layers are: 5\,cm of electroformed copper, 5\,cm of commercial oxygen-free high conductivity copper and 45\,cm of low background lead. The shield will be enclosed and purged with dry boil-off nitrogen to remove radon. This radon enclosure will be fully covered (nearly 4$\pi$) by a muon veto comprised of plastic scintillators. Finally, the muon veto is enclosed in a 5\,cm layer of borated polyethylene and 25\,cm of polyethylene to absorb neutrons. A schematic of the two cryostats inside of the shield is shown in Fig. \ref{fig:shield}.

Backgrounds in the experiment are reduced in several ways besides constructing the experiment underground to shield it from cosmic rays. All the materials used to construct the \demonstrator (electroformed copper, PTFE, cables, LMFEs, etc.) have been screened to have very low levels of uranium and thorium and meet the background goals of the experiment. The assembly of the detector units and their manipulation is entirely conducted in a glove box that is purged with boil-off nitrogen to reduce any possible contamination from radon. The intrinsically good energy resolution of germanium detectors further helps to mitigate backgrounds by enhancing the expected peak relative to the two neutrino mode of double beta-decay. The choice of PPC detectors allows backgrounds associated with multiple scatters (such as high energy gamma rays and $^{68}$Ge decays) to be strongly rejected. A very detailed simulation of the expected backgrounds based on assays and the characterization of the detectors has been performed and predicts a performance in line with the stated goals.

\subsection{Schedule and status}
The \mjd experiment is proceeding in three phases. In the first phase, a prototype cryostat made of commercial OFHC copper is used to deploy two strings of natural (non-enriched) germanium detectors in a partially assembled shield. The goal of this prototype is to demonstrate the integration of the various components (detectors, vacuum, cooling, shielding, data acquisition). The first string of detectors was installed in the prototype in the summer 2013 and tests are currently ongoing (as of October 2013). In the second phase, the first module made from electroformed copper will be populated with a mix of natural and enriched detectors and run inside the completed shield alongside the prototype cryostat. This second phase is anticipated to begin commissioning in the first quarter of 2014. Finally, in the third phase, a second cryostat made of electroformed copper and containing only enriched detectors will replace the prototype cryostat in the shield. Approximately 2700\,kg (85\%) of the electroformed copper have been produced so far, with all parts required for the first module in hand.

The collaboration has also started to produce and test PPC detectors made from germanium enriched in \germaniumsp. The material was enriched by Electrochemical Plant (ECP) in Russia and delivered to the collaboration in the form of germanium oxide powder. The oxide powder was transported by boat and ground transport in a shielded container to a shallow underground site near Oak Ridge, TN, USA. The oxide is reduced and zone refined to electronic grade material by ESI in Oak Ridge. The electronic grade material is then provided to AMETEK/ORTEC which has been contracted by the collaboration to fabricate the diodes. Care is taken to minimize the exposure of the enriched material to cosmic rays, thus reducing comsogenic activation of isotopes that could contribute backgrounds (in particular, $^{60}$Co and $^{68}$Ge). The fabricated detectors are tested briefly at ORTEC and, if accepted by the collaboration, immediately transported to SURF for further detailed characterization. As of October 2013, the collaboration has accepted approximately half of the 30\,kg of enriched detectors, which are currently underground at SURF and ready for deployment.

\section{Other Physics and MALBEK}
The use of PPC detectors and very low levels of background allow the \mjd experiment to perform searches for other rare events in addition to the search for \nbb decay. PPC detectors have a very low energy threshold because the small capacitance of the point contact electrode results in very low levels of electronic noise. This low threshold allows searches for dark matter \cite{Friedland:2012fa}, solar axions \cite{Creswick:1997pg}, violation of the Pauli Exclusion Principle \cite{Elliott:2012pep}, and coherent elastic neutrino nucleus scattering \cite{Wong:2005vg}.

The {\sc Majorana} Collaboration is currently operating the MALBEK detector \cite{Aalseth:2010zv} at the Kimbalton Underground Research Facility (KURF) in Virginia, USA, at 1450 meters of water equivalent shielding. This Broad Energy Germanium (BEGe) detector was manufactured by CANBERRA and is operated in a low background cryostat with shielding and an active muon veto. MALBEK is being used by the collaboration to optimize the background simulation software, to test the data acquisition system for the \demonstratorsp, and to perform a search for dark matter and solar axions. Fig. \ref{fig:MALBEK} shows an example of the excellent agreement between the detailed simulation of the backgrounds and the data taken with the detector. 

\begin{ltxfigure}[ht]
\centering
  \subfloat[Fit to background spectrum]{\includegraphics[width=0.45\textwidth]{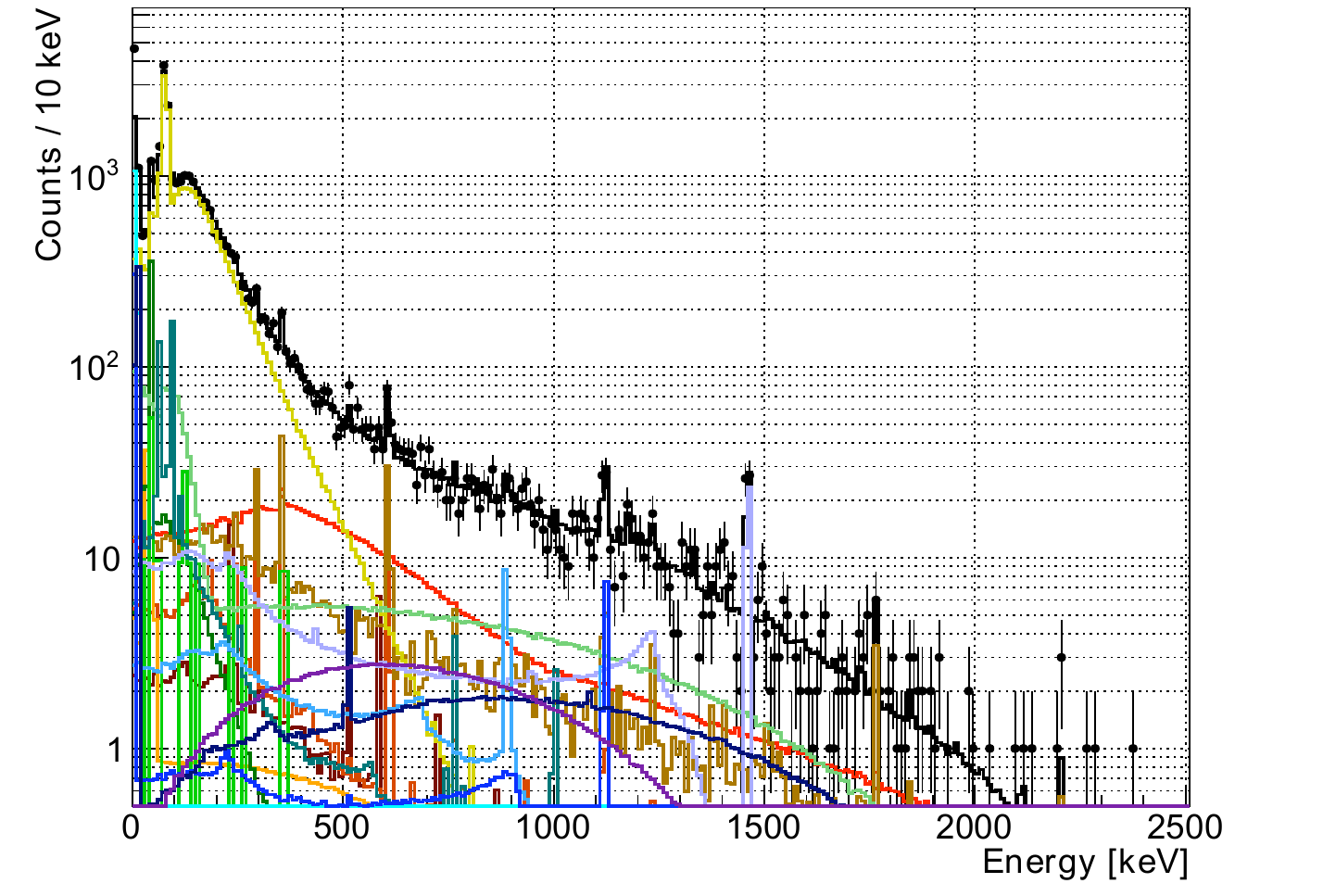}}  
  \subfloat[Components in fit]{\includegraphics[width=0.45\textwidth]{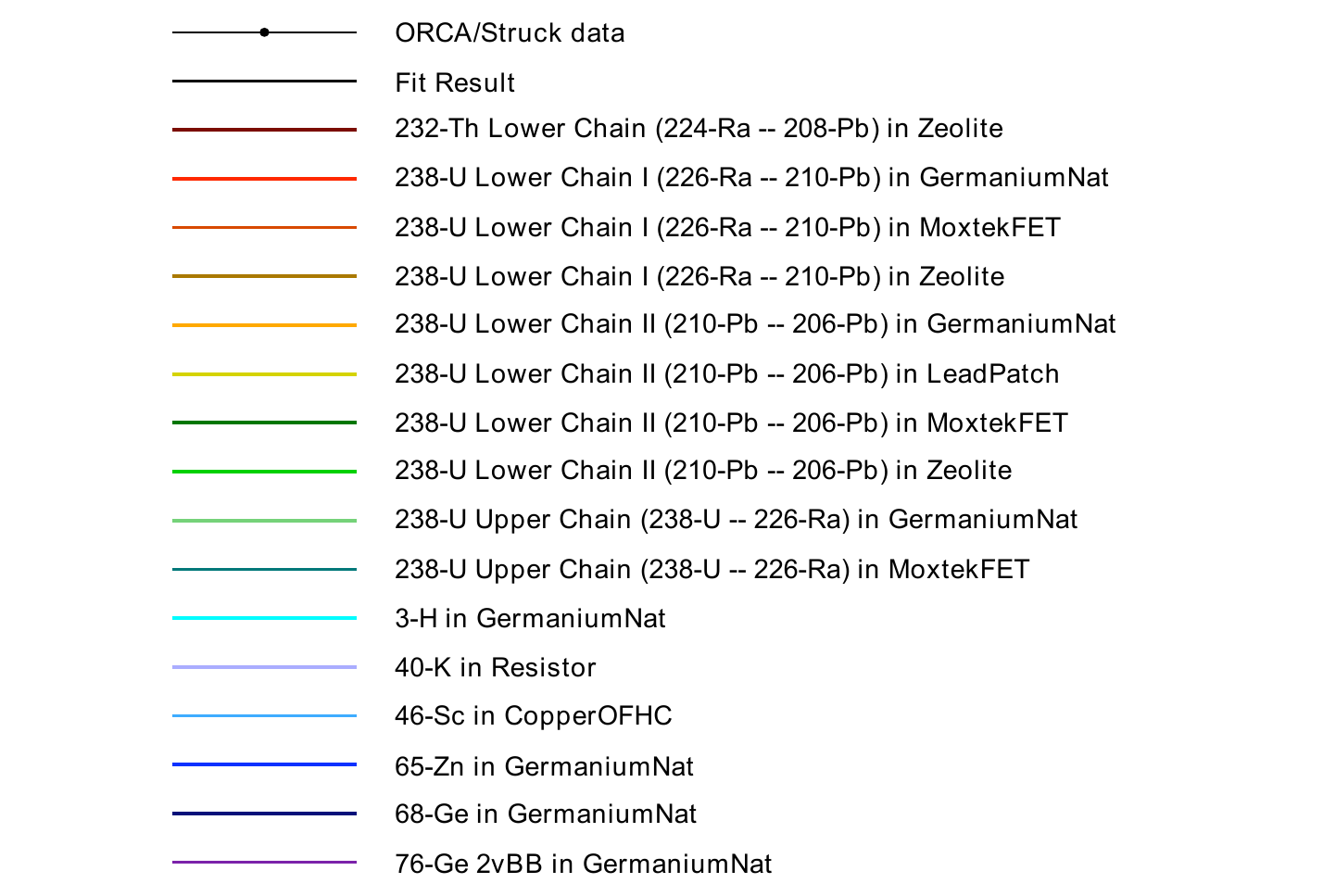}}    
  \caption{\label{fig:MALBEK}Comparison of data and the simulated background in the MALBEK detector \cite{Schubert:2012}. Color figure available online.}
\end{ltxfigure}

\section{Conclusion}
The \mjd experiment is currently under construction at the Sanford Underground Research Facility in South Dakota, USA. First data using the enriched detectors are expected in 2014. The main goals of the experiment are to search for neutrinoless double beta-decay with an exposure of the order of 100\,kgy and to demonstrate a technology that has low enough backgrounds to justify the construction of a tonne scale experiment. The \demonstrator will also have a rich research program in dark matter detection, solar axions, and other fields that complement the search for neutrinoless double beta-decay.

\bibliographystyle{aipproc}   

\bibliography{MJReferences}

\end{document}